\documentstyle[12pt]{article}
\input epsf

\begin{document}
\baselineskip=15pt

\newcommand{\be}{\begin{equation}}
\newcommand{\ee}{\end{equation}}
\newcommand{\bq}{\begin{eqnarray}}
\newcommand{\eq}{\end{eqnarray}}
\newcommand{\x}{{\bf x}}
\newcommand{\y}{{\bf y}}
\newcommand{\one}{\hbox{\rm 1\kern-.27em I}}
\newcommand{\p}{\varphi}
\newcommand{\Sc}{Schr\"odinger\,}
\newcommand{\del}{\nabla}
\begin{titlepage}
\vskip1in
\begin{center}
{\large The Lagrangian origin of MHV rules } 
\end{center}
\vskip1in
\begin{center}
{\large

Paul Mansfield 

Department of Mathematical Sciences

University of Durham

South Road

Durham, DH1 3LE, England}

{\it P.R.W.Mansfield@durham.ac.uk}
\end{center}
\vskip1in
\begin{abstract}

\noindent 
We construct a canonical transformation that takes the usual Yang-Mills action into one whose Feynman diagram expansion generates the MHV rules. The off-shell continuation appears as a natural consequence of using light-front quantisation surfaces. The construction extends to include massless fermions.
\end{abstract}

\end{titlepage}

\section{\bf Introduction}

One of the most remarkable aspects of recent developments in the
perturbative approach to gauge theories 
is the demonstration that known results for scattering amplitudes, at least at tree-level
\cite{Witten}-\cite{CSW}
and low order in the loop expansion \cite{bill}-\cite{billf},
can be constructed by sewing together simpler scattering amplitudes.
This appears to offer an alternative to the usual Feynman diagram expansion.
A special r\^ole is played by the
maximally helicity violating amplitudes (MHV) which describe the tree-level scattering of 
$n$ gluons, with $n-2$ of positive helicity and 2 of negative helicity
(when all gluons are assigned outgoing momenta.) Any tree-level amplitude,
${\cal A}_n$,
can be decomposed into a sum over colour-ordered partial amplitudes, $A_n$,
multiplied by
a momentum conserving delta function, and a trace over a product of colour
matrices, 
$$ 
{\cal A}_n=\sum_{\sigma} {\rm tr}\, (T^{R_{\sigma(1)}}..T^{R_{\sigma(n)}})\,i (2\pi)^4\,\delta^4(
p^1+..+p^n)\,A_n^\sigma\,,
$$
where the sum extends over distinct cyclic orderings of the gluons, $\sigma$.
For an MHV amplitude the partial amplitude has the simple form \cite{parke}-\cite{ber}
$$
A=g^{n-2}{\langle \lambda_r,\,\lambda_s\rangle\over \prod_{j=1}^n \langle \lambda_j,\,\lambda_{j+1}\rangle}
$$
where the gluons with negative helicity are labelled by $r$ and $s$ and $g$ is the coupling. The gluons are on-shell, and if the $j$-th has four-momentum 
components $(p^\mu)$
with respect to the Cartesian co-ordinates $(t,\,x^1,\,x^2,\,x^3)$,
then $\lambda_j$ is the two-component spinor such that $\lambda_j\,\tilde\lambda_j=
p^t+\sum\sigma^ip^i\equiv Q(p)$, where $\sigma^i$ are the Pauli matrices
and $\tilde\lambda=\lambda^\dagger$ for positive energies and $\tilde\lambda=-\lambda^\dagger$ for negative ones, and $\langle \lambda_j,\,\lambda_{k}\rangle =\lambda_j^T \,i\sigma^2\,\lambda_k$, where $T$ denotes tranposition.

In the MHV rules the partial amplitudes
replace the vertices of the usual Feynman diagrams, (see \cite{Nigel} and references therein for recent developments in this approach),
and these are glued together using scalar propagators that contract fields
of opposite helicity. Internal lines are off-shell, so that a prescription
is need to continue the MHV amplitude. 
The investigation of these rules has been largely empirical in that a dual string theory picture first inspired the conjecture of rules 
for combining amplitudes, first at tree-level and then at loop level, and these
conjectures have been tested against known results. Recently the rules have been derived from a twistor-space action, \cite{Mason}.
In this paper we will 
derive the MHV rules directly from Yang-Mills theory by constructing a 
canonical transformation that maps between the two. This makes the details of the 
MHV rules such as the off-shell continuation transparent, as well as taking a step towards the systematic development of the quantum theory via the loop expansion.

\section{The transformation}

It is well known that the light-front quantisation of Yang-Mills theory leads to a simple formulation in terms of physical degrees of freedom, so we begin
by writing the covariant action
$$
S={1\over 2 g^2}\int dt\,dx^1\,dx^2\,dx^3 \,{\rm tr}\, \left(F^{\lambda \rho}\,F_{\lambda \rho}\right),$$
where
$$ F_{\lambda \rho}=[{\cal D}_\lambda,\,{\cal D}_\rho],
\quad {\cal D}=\partial+A,\quad A=A^RT^R,$$
$$ [T^R,T^S]=f^{RSP}T^P,
\quad {\rm tr}\,\left(T^R\,T^S\right)=-{\delta^{RS}\over 2},
$$
in terms of variables appropriate to quantisation
surfaces of constant $\mu\cdot x$, where $\mu$ is a null-vector. We will
use space-time co-ordinates related to the Cartesian co-ordinates
$(t,x^1,x^2,x^3)$ for which $(\mu)=(1,0,0,1)$ by
\be
x^0=t-x^3,\quad x^{\bar o}=t+x^3,\quad z=x^1+ix^2, \quad {\bar z}=x^1-ix^2\,,
\label{coord}
\ee
so that the invariant interval is $ds^2=dx^0\,dx^{\bar o}-dz\,d{\bar z}$. It is natural to impose the gauge $\mu\cdot A=0\Rightarrow A_{\bar 0}=0$, which leads 
to the field independent Faddeev-Popov determinant ${\rm Det} \,\partial_{\bar 0}$. This 
eliminates one unphysical degree of freedom, we can make the other explicit by
defining 
\be A_L=A_0-\partial_{\bar 0}^{-1}\left(\partial_{\bar z} A_z+\partial_z A_{\bar z}
\right)\,. \label{div}
\ee
This corresponds to expanding the gauge fixed field on the quantisation surface
as 
\be A^\rho=\mu^\rho \,A_L+A_+^\rho+A_-^\rho\,,\label{decomp}\ee
where $A_\pm$ contain the physical positive and negative
helicity components with polarisation vectors $E_\pm$ associated with the Fourier expansions
$$A_\pm=
\int {d^4p} \,\delta (p\cdot p){E}_\pm(p)\left(\,a({\bf p},x^0)_\alpha^r\,e^{-ip\cdot x}+
b({\bf p},x^0)^r_\alpha\,e^{ip\cdot x}\right)\,,
$$
so that $p$ is on-shell with positive energy. Because of the arbitrary $x^0$ dependence included in the coefficients, $a$ and $b$, this places no restriction on $A$ other than the gauge-condition, and that it have a Fourier integral.
If these coefficients were independent of $x^0$ then $A$ would be on-shell, but we do not assume this. The $\{E_r\}$ are most conveniently expressed as quaternions
$$E_+(p)={\mu_s\tilde\lambda\over \langle\,\mu_s\,,\lambda\,\rangle},
\quad E_-( p)={\lambda\tilde\mu_s\over[\,\lambda,\,\mu_s]}\,,
$$
where
$[\,\lambda,\,\mu_s]=\tilde\lambda i\sigma^2\tilde\mu_s^T$,
and $\mu_s$ is a 2-spinor related to the null-vector $\mu$ by
$Q(\mu)=\mu_s\,\tilde\mu_s$, for example $\mu_s=(\sqrt 2,0)^T$. The polarisations satisfy $p\cdot E_\pm (p)=0$ which leads to (\ref{div}).
Now in the co-ordinates (\ref{coord}) 
$$
\mu_z=\mu_{\bar z}=0,\quad E_+(p)_{\bar z}=E_-(p)_{z}=0
$$
so that in (\ref{decomp}) only the positive helicity field $A_+$ contributes to
$A_z$ whilst only the negative helicity field $A_-$ contributes to $A_{\bar z}$.

In these variables the action becomes, 
$S={1\over g^2}\int dx^0\,dx^{\bar 0}\,dz\,d{\bar z}\,\left(
{\cal L}_2+{\cal L}_3+{\cal L}_4\right)
$ with
$$
{\cal L}_2={\rm tr}\, \left( A_z\,\partial^2\,A_{\bar z}-(\partial_{\bar 0}A_L)^2
\right)\,,
$$
\newpage
$$
{\cal L}_3={\rm tr}\, \Big((\partial_{\bar 0}A_z)\,[A_{\bar z},A_L]
+(\partial_{\bar 0}A_{\bar z})\,[A_{z},A_L]$$
$$+4(\partial_{\bar 0}A_z)\,[A_{\bar z},\partial_{\bar 0}^{-1}\partial_z A_{\bar z}]+4(\partial_{\bar 0}A_{\bar z})\,[A_{ z},\partial_{\bar 0}^{-1}\partial_{\bar z} A_{ z}]\Big)\,,
$$
$$
{\cal L}_4={\rm tr}\, \left( A_z\,A_{\bar z}\,[A_{\bar z},\,A_z]\right)\,.$$
This is quadratic in $A_L$, so we can integrate out this degree of freedom.
Equivalently,
if we were taking a Hamiltonian point of view, rather than a Lagrangian one, then we 
would observe that the equation of motion of $A_L$ does not involve the `time' derivative
$\partial _0$ appropriate to the constant-$x^0$ quantisation surfaces, $\Sigma$,
and so this variable should be eliminated via its equation of motion. The resulting action takes the particularly 
compact form
$$S_L={4\over g^2}\int dx^0\,d^3{\bf x}\,{\rm tr}\, \left(
A_z\,\partial_0\partial_{\bar 0}\,A_{\bar z}-
[{\cal D}_{\bar z},\,\partial_{\bar 0} A_z]\,\partial_{\bar 0}^{-2}\,
[{\cal D}_{z},\,\partial_{\bar 0} A_{\bar z}]\right)
\,,$$
where $d^3{\bf x}=dx^{\bar 0}\,dz\,d{\bar z}$ and bold-face type refers to position on constant-$x^0$ surfaces.
Writing out the gauge-covariant derivatives gives the light-front Lagrangian in the form
$L_2+{L}^{++-}+{L}^{--+}+{L}^{--++}$ with
$$
L_2[A]={4\over g^2}\int_\Sigma d^3{\bf x}\,{\rm tr}\, \left(A_z\,(\partial_0\partial_{\bar 0}-\partial_z\partial_{\bar z})\,A_{\bar z}\right)\,,
$$
$${L}^{++-}[A]={4\over g^2}\int_\Sigma d^3{\bf x}\,{\rm tr}\, \left(-
({\partial}_{\bar z}\partial_{\bar 0}^{-1} A_z)\,
[{A}_{z},\,\partial_{\bar 0} A_{\bar z}]\right)\,,$$
$$
{L}^{--+}[A]={4\over g^2}\int_\Sigma d^3{\bf x}\,{\rm tr}\, \left(-
[{A}_{\bar z},\,\partial_{\bar 0} A_z]\,
({\partial}_{z}\partial_{\bar 0}^{-1} A_{\bar z})\right)\,,$$
$$
{L}^{--++}[A]={4\over g^2}\int_\Sigma d^3{\bf x}\,{\rm tr}\, \left(-
[{A}_{\bar z},\,\partial_{\bar 0} A_z]\,\partial_{\bar 0}^{-2}\,
[{A}_{z},\,\partial_{\bar 0} A_{\bar z}]\right)\,,$$
In the Feynman diagram expansion $L_2$ gives a scalar type propagator
$\propto 1/p^2$ contracting the positive and negative helicity fields
$A_z$ and $A_{\bar z}$ contained in the vertices ${L}^{++-}$, 
${L}^{--+}$, and ${L}^{--++}$ which are labelled by their helicity content. The 
resulting diagrams differ significantly from the MHV rules because they involve 
the vertex ${L}^{++-}$ which has only one negative helicity and the
higher order vertices corresponding to the maximal helicity violating amplitudes
themselves are absent.

To change the action into one that generates the MHV rules we look for a 
transformation that effectively
eliminates the vertex ${L}^{++-}$ at the same time generating the missing MHV vertices. We will require that the transformation be canonical because in light-front quantisation the momentum canonically conjugate to $A_z$, 
$\Pi_z$, is (up to a constant)
${\partial_{\bar 0} A_{\bar z}}$ so that the functional integral measure 
obtained as the product over space-time of
$dA_z(x)\,dA_{\bar z}(x)$ differs from 
the product of $dA_z(x)\,d\Pi_{z}(x)$
by the field independent factor ${\rm Det}(\partial_{\bar 0})$ and so is invariant
under canonical transformations. So we look for new fields $B_\pm (x)$ such that $B_+$ is a functional of
$A_z$ on the quantisation surface, (but not $A_{\bar z}$), $B_+=B_+[A_z]$,
and 
\be  \partial_{\bar 0} A_{\bar z}(x^0,\,{\bf y})=
\int_{\Sigma}d^3{\bf x}\,{\delta B_+(x^0,\,{\bf x})\over\delta A_z(x^0,\,{\bf y})}\partial_{\bar 0}B_-(x^0,\,{\bf x})\,.\label{can}
\ee
We choose the transformation to ensure that
$$
L_2[A]+{L}^{++-}[A]=L_2[B]\,,
$$
so when we express the free part of the action and the unwanted vertex in terms of the new fields we obtain just a free action.
Explicitly this requires
$$
\int_{\Sigma}d^3{\bf y}\,{\rm tr}\, \left(\left(\{\partial_0-\omega\}A_z-[A_z,\partial_{\bar z}\partial_{\bar 0}^{-1}A_z]\right)|_{x^0,{\bf y}}
\,{\delta B_+(x^0,\,{\bf x})\over\delta A_z(x^0,\,{\bf y})}\partial_{\bar 0}B_-(x^0,\,{\bf x})\right)$$
$$
={\rm tr}\, \left(\{\partial_0-\omega\}B_+\partial_{\bar 0}B_-\right)|_{x^0,{\bf x}}\,.
$$
where we have introduced the operator $\omega({\bf x})=\partial_z\partial_{\bar z}/\partial_{\bar 0}$.
The terms in $\partial_0 A_z$ and $\partial_0 B_+$ are automatically equal
provided that $B_+$ depends on $x^0$ only implicitly through $A_z$, so that
$B_+$ just has to satisfy
$$
\int_{\Sigma}d^3{\bf y}\, [{\cal D}_z,{\partial_{\bar z}\partial_{\bar 0}^{-1}}A_z]|_{x^0,{\bf y}}
\,{\delta B_+(x^0,\,{\bf x})\over\delta A_z(x^0,\,{\bf y})}
=\omega\,B_+(x^0,\,{\bf x})\,.
$$
This is readily solved as a power series in $A_z$ so that 
$$
B_+^R(x^0,\,{\bf x})=\sum_{n=1}^\infty \int_{\Sigma} d^3{\bf y}_1..d^3{\bf y}_n \Gamma_n^{RP_1..P_n}({\bf x},{\bf y}_1..{\bf y}_n)\,A_z^{P_1}(x^0,\,{\bf y}_1)..A_z^{P_n}(x^0,\,{\bf y}_n)\,
$$
where the functions $\Gamma_n$ are independent of $x^0$ and are constructed iteratively from
$$
\Gamma_1^{RP_1}({\bf x},{\bf y}_1)=\delta^{RP_1}\,\delta^3({\bf x}-{\bf y}_1)\,,\quad
\Gamma_n^{RP_1..P_n}({\bf x},{\bf y}_1..{\bf y}_n)=$$
\be S\,{1\over \omega({\bf x})+\omega({\bf y}_1)+..+\omega({\bf y}_n)}
f^{P_1P_2 P}\left({\partial_{\bar z}\over\partial_{\bar 0}}\,\delta({\bf y}_1-{\bf y}_2)\right)
\,\Gamma_{n-1}^{RPP_3..P_n}({\bf x},{\bf y}_2..{\bf y}_n)\,.
\label{sol} \ee
$S$ is the instruction to symmetrise over the pairs of indices attached to the 
$A_z$ fields, $P_1,\,{\bf y}_1..P_n,\,{\bf y}_n$.
The inverse of the transformation gives $A_z$ on $\Sigma$ as a power series in $B_+$
of the form
\be
A_z^R(x^0,\,{\bf x})=\sum_{n=0}^\infty \int_{\Sigma} d^3{\bf y}_1..d^3{\bf y}_n \Upsilon_n^{RP_1..P_n}({\bf x},{\bf y}_1..{\bf y}_n)\,B_+^{P_1}(x^0,\,{\bf y}_1)..B_+^{P_n}(x^0,\,{\bf y}_n)\,,\label{sol2} 
\ee
with the $\Upsilon$ computable from $\Gamma$
and from this we obtain $A_{\bar z}$ as a power series using (\ref{can}) 
$$
A_{\bar z}^R(x^0,\,{\bf x})=$$
\be{1\over\partial_{\bar 0}}
\sum_{n=1}^\infty \int_{\Sigma} d^3{\bf y}_1..d^3{\bf y}_n n\,\Xi_n^{RP_1..P_n}({\bf x},{\bf y}_1..{\bf y}_n)\,B_+^{P_1}(x^0,\,{\bf y}_1)..B_+^{P_{n-1}}(x^0,\,{\bf y}_{n-1})\,\partial_{\bar 0}B_-^{P_n}(x^0,\,{\bf y}_n)\,.
\label{mom}
\ee
The important point is that this last expression is linear in $\partial_{\bar 0}B_-$ So that when the remaining part of the Lagrangian, ${L}^{--+}[A]+
{L}^{--++}[A]$, is expressed in terms of $B_\pm$ the result is an infinite series in $B_+$ but is only quadratic in $B_-$. We write this as $V^{--+}[B]+
{V}^{--++}[B]+{V}^{--+++}[B]+..$. The vertices are labelled by their 
helicity content in terms of the positive helicity $B_+$ field and negative helicity
$B_-$ field, and are local in the light-front `time', $x^0$,
\newpage
$$
V^{--+..+}=\int_{\Sigma} d^3{\bf y}_1..d^3{\bf y}_n \,{\tilde V}^{{P_1..P_n}}({\bf y}_1,..,{\bf y}_n)\,B_-^{P_1}(x^0,\,{\bf y}_1)\,\,B_-^{P_2}(x^0,\,{\bf y}_2)\,
\times
$$
\be B_+^{P_{3}}(x^0,\,{\bf y}_{3})\,...\,B_+^{P_n}(x^0,\,{\bf y}_n)\,.\label{vert}
\ee

In principle we could obtain explicit expressions for the vertices $V^{--+..+}$ from the transformation (\ref{sol}), but we will take a different path so that we will not need the detailed form of the transformation; only its existence and general properties will be used. We will construct the off-shell Lagrangian from a knowledge of on-shell tree-level scattering. We do not include loops because the Lagrangian itself is a classical object. Consider 
calculating an MHV amplitude with $n$ on-shell gluons from the
Feynman diagram expansion of the transformed action
$$S_L={4\over g^2}\int dx^0\, \left(
L_2[B]+V^{--+}[B]+
{V}^{--++}[B]+..+{V}^{--+..+}[B]+..\right)\,.
$$
The LSZ procedure gives the amplitude in terms of the momentum space Green function for $n-2$ suitably normalised $A_z$ fields and two $A_{\bar z}$ fields by cancelling each external leg
using a factor $p^2$ and then taking each momentum on-shell, $p^2\rightarrow 0$.
The equivalence theorem for S-matrix elements allows us to use Green functions for the $B_\pm$ fields instead of the $A_z$ and $A_{\bar z}$, provided we include a multiplicative wave-function renormalisation.
But in calculating the MHV amplitude we are working at tree-level and so obtain identical results using the $B$ fields or the $A$ fields because to leading order in the expansions (\ref{sol2}) and (\ref{mom}) they are the same, and the higher order terms are annihilated by the on-shell $p^2$ factors that cancel externel legs since we don't include loops. The MHV amplitude is therefore the sum of on-shell tree-level Feynman diagrams with $n-2$ external $B_+$ legs and two external $B_-$ legs, with the propagators for the external legs cancelled. Since the propagator contracts $B_+$ fields with $B_-$ fields in pairs there is only one vertex that can contribute to any given MHV amplitude, namely the vertex with the same helicity assignment, (although amplitudes that are not MHV will be made up of contractions of more than one vertex.) So the MHV amplitude is simply the vertex evaluated on-shell. This provides useful,
but limited information about the Lagrangian.
Of course we really need the vertices evaluated for arbitrary field configurations, not just those that are on-shell, but the general properties of the canonical transformation will enable us to extract these and at the same time
shed light on the origin of the off-shell continuation proposed in \cite{Witten}.
Explicitly we replace the $B_+ (x^0,{\bf y})$ fields in (\ref{vert}) by $g\,T^R\,E_z^+e^{ip\cdot y}$ with $p\cdot y=p_0x^0+p_{\bar 0}y^{\bar 0}+p_zy^z+p_{\bar z}y^{\bar z}
\equiv p_0x^0+{\bf p}\cdot {\bf y}$ and 
$B_- (x^0,{\bf y})$ fields by $g\,T^R\,E_{\bar z}^-e^{ip\cdot y}$ (both with $p^2=0$) to obtain the MHV amplitude as 

$$
{4g^{n-2}}\int dx^0\,d^3{\bf y}_1..d^3{\bf y}_n \,{\tilde V}^{{R_1..R_n}}({\bf y}_1,..,{\bf y}_n)\, e^{i\sum_{1}^n (p_{0}^jx^0+{\bf p}^j\cdot {\bf y}_j)}
E_z^+({\bf p}^1)..E_z^+({\bf p}^n)\,E_{\bar z}^-({\bf p}^r)\,E_{\bar z}^-({\bf p}^s)
$$
$$
=\sum_{\sigma} i (2\pi)^4\,\delta^4(p^{1}+..+p^{n})\,g^{n-2}\, {\rm tr}\,(T^{R_{\sigma(1)}}..T^{R_{\sigma(n)}})\,{\langle \lambda_r,\,\lambda_s\rangle^4\over \prod_{j=1}^n \langle \lambda_{\sigma(j)},\,\lambda_{\sigma(j+1)}\rangle}
$$
$\tilde V$ is independent of $x^0$ so this integration yields a $\delta$-function
giving an expression for the Fourier transform of $\tilde V$:
\newpage
$$
\delta\left(p_{0}^1+..+p_{0}^n\right)\, \int d^3{\bf y}_1..d^3{\bf y}_n \,{\tilde V}^{{R_1..R_n}}({\bf y}_1,..,{\bf y}_n)\, e^{i\sum_1^n {\bf p}^j\cdot {\bf y}_j}=
$$
$$
\delta(p_{0}^1+..+p_{0}^n) \,\sum_{\sigma} i\pi^3\,\delta^3({\bf p}^{1}+..+{\bf p}^{n})\,{{\rm tr}\, (T^{R_{\sigma(1)}}..T^{R_{\sigma(n)}})\over E_z^+({\bf p}^1)..E_z^+({\bf p}^n)\,E_{\bar z}^-({\bf p}^r)\,E_{\bar z}^-({\bf p}^s)}\,{\langle \lambda_r,\,\lambda_s\rangle^4\over \prod_{j=1}^n \langle \lambda_{\sigma(j)},\,\lambda_{\sigma(j+1)}\rangle}
$$
The Fourier transform of $\tilde V$ specifies the vertex completely so that
once it is known we can compute (\ref{vert}) for arbitrary, off-shell field
configurations. It would appear to be a simple matter to cancel the first 
$\delta$-function to obtain what we need:
$$
\int d^3{\bf y}_1..d^3{\bf y}_n \,{\tilde V}^{{R_1..R_n}}({\bf y}_1,..,{\bf y}_n)\, e^{i\sum_1^n {\bf p}^j\cdot {\bf y}_j}=
$$
\be
\sum_{\sigma} i\pi^3\,\delta^3({\bf p}^{1}+..+{\bf p}^{n})\,{{\rm tr}\, (T^{R_{\sigma(1)}}..T^{R_{\sigma(n)}})\over E_z^+({\bf p}^1)..E_z^+({\bf p}^n)\,E_{\bar z}^-({\bf p}^r)\,E_{\bar z}^-({\bf p}^s)}\,{\langle \lambda_r,\,\lambda_s\rangle^4\over \prod_{j=1}^n \langle \lambda_{\sigma(j)},\,\lambda_{\sigma(j+1)}\rangle}\,,\label{solft}
\ee
however there is the possibility of missing a term that vanishes on the support of the cancelled $\delta$-function. We now appeal to analyticity to show that such a term is absent.

Firstly observe that the vertices, (\ref{vert}), are constructed from 
\be
L^{--+}[A]+L^{--++}[A]=
{4\over g^2}\int_\Sigma d^3{\bf x}\,{\rm tr}\, \left(-
[{A}_{\bar z},\,\partial_{\bar 0} A_z]\,\partial_{\bar 0}^{-2}\,
[{\cal D}_{z},\,\partial_{\bar 0} A_{\bar z}]\right)\,,\label{orig}
\ee
which is written without the use of $\partial_{\bar z}$. As we will see,
this implies that
the vertices, (\ref{vert}), inherit this property. The canonical transformation
(\ref{sol}) does involve $\partial_{\bar z}$, both explicitly, and in the
operator $\omega({\bf x})=\partial_z\partial_{\bar z}/\partial_{\bar 0}$. This
dependence cancels for $n=2$, but not for larger $n$, so we need to study the effect on the transformation of varying $\partial_{\bar z}A_z$. Now we constructed the transformation so that
$$
L_2[B]=L_2[A]+{L}^{++-}[A]=
{4\over g^2}\int_\Sigma d^3{\bf x}\,{\rm tr}\, \left(A_z\,\partial_0\partial_{\bar 0}\,A_{\bar z}+
{\partial}_{\bar z}A_z\,\partial_{\bar 0}^{-1} \,
[{\cal D}_{z},\,\partial_{\bar 0} A_{\bar z}]\right)\,.$$
This is almost invariant under the homogeneous part of a gauge transformation with a gauge-parameter $\theta$ that depends only on $\bar z$, 
\be
\delta_\theta\, A_z=[A_z,\theta(\bar z)],\quad\delta_\theta\, A_{\bar z}=[A_{\bar z},\theta(\bar z)]\,\label{ahol}
\ee
but fails to be so because of the second term in the transformation of
${\partial}_{\bar z}A_z$ 
$$
\delta_\theta \,{\partial}_{\bar z}A_z=[{\partial}_{\bar z}A_z,\theta(\bar z)]
+[A_z,{\partial}_{\bar z}\theta(\bar z)]\,. $$
Consequently the
change in $L_2[A]+{L}^{++-}[A]$ is the same as
if we only vary ${\partial}_{\bar z}A_z$ by this second term, $\delta\,
{\partial}_{\bar z}A_z=[A_z,{\partial}_{\bar z}\theta(\bar z)]$
and leave 
everything else alone. So the effect on the canonical transformation of varying 
$\partial_{\bar z}A_z$ is equivalent to a transformation of the form of
(\ref{ahol}), but (\ref{orig}) is manifestly invariant under such a change, so the vertices, (\ref{vert}), cannot contain $\partial_{\bar z}$.
This explains the absence from (\ref{solft}) of any term that vanishes on the support of 
$\delta( p_{0}^1+..+p_{0}^n)$ for on-shell $p_0=p_zp_{\bar z}/p_{\bar 0}$,
because
for $n>3$ any such term depends on $p_{\bar z}$. ($n=3$ is a special case because
using 3-momentum conservation
$$
p_{0}^1+p_{0}^2+p_{0}^3={p_z^1p_{\bar z}^1\over p^1_{\bar 0}}+
{p_z^2p_{\bar z}^2\over p^2_{\bar 0}}-{(p_z^1+p_z^2)(p_{\bar z}^1+p_{\bar z}^2)\over p^1_{\bar 0}+p^2_{\bar 0}}={|p^1_zp^2_{\bar 0}-p^2_zp^1_{\bar 0}|^2
\over (p_{\bar 0}^1+p_{\bar 0}^2)p_{\bar 0}^1p_{\bar 0}^2}
$$
so that $p^1_zp^2_{\bar 0}-p^2_zp^1_{\bar 0}$ vanishes on the support
despite being independent of $p_{\bar z}^1$ and $p_{\bar z}^2$. However (\ref{solft})
still holds for this case as can be checked by explicit computation of the
off-shell three-point vertex.)

Consistency requires that, apart from the $\delta$-function, the right-hand-side of (\ref{solft}) should also be independent of the $p_{\bar z}^i$. 
Up to a choice of phase the $\lambda_j$
can be written entirely in terms of $p_{\bar 0}$ and $p_{z}$, and the arbitrariness this
choice of phase is cancelled by the contributions of the polarisation vectors
$E^\pm$ to the denominator, enabling us to take, for example,
$\lambda=(-p_z\sqrt{2}/\sqrt p_{\bar 0},\sqrt{2p_{\bar 0}})^T$ and $E_z^+=-1/2$,
so (\ref{solft}) is indeed independent of the $p_{\bar z}^i$.

This identification of the vertices explains the off-shell continuation used in \cite{Witten}. The 
vertices (\ref{vert}) require $\tilde V$ to be integrated against $B$ fields
that are not constrained to be on-shell, i.e. the vertex will include Fourier components $\int d^4x \,B
(x)\, e^{ip\cdot x} \equiv \tilde B(p_0,{\bf p})$ for which $p^2\neq 0$.
But (\ref{solft}) shows that
the vertex is essentially the MHV amplitude built out of on-shell momenta whose
components within the quantisation surface coincide with the $\bf p$ of 
$B(p_0,{\bf p})$, but with the 0-component fixed by the mass-shell condition. The on-shell momentum constructed from an off-shell momentum 
with the same $\bf p$ part can be expressed as $p-\mu \,p\cdot p/(2\,p\cdot\mu)$.
If $Q(p)$ is the quaternion constructed from $p$,
then
the spinor $\lambda$ to be used in the MHV amplitude satisfies
$$Q(p)-\mu_s\tilde\mu_s \,p^2/(2\,p\cdot\mu)
=\lambda\,\tilde\lambda \quad\Rightarrow \lambda\propto 
Q(p)\eta\,,
$$
which is the prescription of \cite{Witten} when $\eta$ is chosen so that
$\tilde\mu_s\,\eta=0$, i.e. $\eta\propto (0,1)^T$.

Putting all this together gives the transformed action
$$
S_L={4\over g^2}\Bigg(\int dx^0\,d^3{\bf x}\,{\rm tr}\, \Big(B_+\,(\partial_0\partial_{\bar 0}-\partial_z\partial_{\bar z})\,B_-\Big)$$
$$
+\sum_{n=3}^\infty\sum_{\sigma}
\int dx^0\,d^3{\bf p}^1..d^3{\bf p}^n
{{\rm tr}\, (B_+(x^0,{\bf p}^1)..B_-(x^0,{\bf p}^r)..B_-(x^0,{\bf p}^s)..B_+(x^0,{\bf p}^n))\over E_z^+({\bf p}^1)..E_z^+({\bf p}^n)\,E_{\bar z}^-({\bf p}^r)\,E_{\bar z}^-({\bf p}^s)}$$
$$
\times \,i\pi^3\,\delta^3({\bf p}^{1}+..+{\bf p}^{n})\,{\langle \lambda_r,\,\lambda_s\rangle^4\over \prod_{j=1}^n \langle \lambda_{\sigma(j)},\,\lambda_{\sigma(j+1)}\rangle}\Bigg)\,,
$$
where $B_\pm(x^0,{\bf p})=\int_\Sigma d^3{\bf x}\,B_\pm(x^0,{\bf y})e^{-i{\bf p}\cdot{\bf y}}$. This generates Feynman rules in which the MHV amplitudes
appear as vertices contracted using scalar propagators.

\section{Quarks}

We can extend the transformation to include  massless quarks.
When we associate helicities with outgoing particles they become identified with chirality. If we use the representation of the $\gamma$-matrices:
\be \gamma^t = \left(
\begin{array}{cc}
0& 1\\ 1&  0
\end{array}\right)\,\quad
\gamma^i = \left(
\begin{array}{cc}
0& -\sigma^i\\ \sigma^i&  0
\end{array}\right)\,\quad
\gamma^5 = \left(
\begin{array}{cc}
1& 0\\ 0&  -1
\end{array}\right)\,,
\ee
and denote the spinor components as 
$$\psi=(\alpha_+,\,\beta_+,\beta_-\alpha_-)^T/\sqrt 2,\quad
\bar\psi=(\bar \beta_+,\,\bar\alpha_+,\,\bar\alpha_-,\,\bar\beta_-)/\sqrt 2\,,$$
then the $\pm$ subscripts refer to helicities and the fermionic contribution to the Lagrangian density becomes in light-front variables
$$
{\cal L}_q=i\bar\psi\,\gamma^\rho{\cal D}_\rho\,\psi=
$$
\be
i(\bar\alpha_+{\cal D}_0\alpha_-+\bar\beta_+\partial_{\bar 0}\beta_-+
\bar\beta_+{\cal D}_z\alpha_-+\bar\alpha_+{\cal D}_{\bar z}\beta_-
\bar\alpha_-{\cal D}_0\alpha_++\bar\beta_-\partial_{\bar 0}\beta_+-
\bar\beta_-{\cal D}_z\alpha_+-\bar\alpha_-{\cal D}_{\bar z}\beta_+)
\ee
The $\beta$ variables have no $\partial_0$ derivatives acting on them.
They are not dynamical on the constant-$x^0$ surfaces, just like $A_L$,
so they should be eliminated via their equations of motion too. Also 
(\ref{div}) implies that there are now $\bar\alpha_\pm\alpha_\mp$
contributions to the equation of motion of $A_L$ modifying the bosonic action.
The full Lagrangian becomes
$L_f+L_2+{L}^{++-}+{L}^{--+}+{L}^{--++}$ with $L_2+{L}^{++-}+{L}^{--+}$ as before,
${L}^{--++}$ modified to
$$
{L}^{--++}[A,\,\bar\alpha,\,\alpha]={L}^{--++}[A]-{1\over 2g^2}
\int d^3{\bf x}\left(2j\,\partial_{\bar 0}^{-2}\left([A_{\bar z},\,\partial_{\bar 0}A_z]+[A_{z},\,\partial_{\bar 0}A_{\bar z}]\right)+j\,\partial_{\bar 0}^{-2}j\right)\,,
$$
where
$$j^P=-{ig^2\over 4}\left(\bar\alpha_+\,T^P\alpha_-+\bar\alpha_-\,T^P\alpha_+\right)\,,
$$
and
$$ 
L_f={i\over 4}\int_\Sigma d^3{\bf x}\,\Big((\bar\alpha_+{\partial}_0\alpha_-
+\bar\alpha_-{\partial}_0\alpha_+
+\bar\alpha_+\left({\partial}_{\bar 0}^{-1}(\partial_{\bar z}A_z+\partial_zA_{\bar z})\right)\alpha_-$$
$$+\bar\alpha_-\left({\partial}_{\bar 0}^{-1}(\partial_{\bar z}A_z
+\partial_zA_{\bar z})\right)\alpha_+-\bar\alpha_+{\cal D}_{\bar z}{\partial}_{\bar 0}^{-1}{\cal D}_{z}\alpha_--\bar\alpha_-{\cal D}_{\bar z}{\partial}_{\bar 0}^{-1}{\cal D}_{z}\alpha_+\Big)\,.
$$
This can be decomposed by helicity into $L_f=L_f^{+-}+{L}^{++-}_f+{L}^{--+}_f+{L}^{--++}_f\,.$

We now look for a transformation to new variables $B_\pm,\,\bar\xi_\pm$ and $\xi_\pm$that eliminates the $++-$ vertices whilst
preserving the integration measure ${\cal D} A_z\,{\cal D} A_{\bar z}\,
{\cal D} \bar \alpha_+\,{\cal D} \bar \alpha_-\,{\cal D} \alpha_+\,{\cal D} \alpha_-$.
To fulfil the second requirement we again take the transformation to be canonical,
and since $i\bar\alpha_\mp/4$ is conjugate to $\alpha_\pm$ we take this to have the form
$$B_+=B_+[A_z,\,\xi_+,\,\xi_-]\,,\quad \xi_\pm(x^0,{\bf x})=\int_{\Sigma}d^3{\bf y}\,R({\bf x}, {\bf y})\,
\alpha_\pm(x^0,{\bf y})\,,
$$
$$
  \partial_{\bar 0} A_{\bar z}(x^0,\,{\bf y})=
\int_{\Sigma}d^3{\bf x}\,{\delta B_+(x^0,\,{\bf x})\over\delta A_z(x^0,\,{\bf y})}\partial_{\bar 0}B_-(x^0,\,{\bf x})+$$
$${i2g^2}\int_{\Sigma}d^3{\bf x}\,d^3{\bf x}'\,\left(
\bar\xi_+(x^0,{\bf x})\,{\delta R({\bf x},{\bf x}')\over\delta A_z(x^0,\,{\bf y})}\,\alpha_-(x^0,{\bf x}')+\bar\xi_-(x^0,{\bf x})\,{\delta R({\bf x},{\bf x}')\over\delta A_z(x^0,\,{\bf y})}\,\alpha_+(x^0,{\bf x}')\right)\,,
$$
\be
\bar\alpha_\pm(x^0,{\bf x})=\int_{\Sigma}d^3{\bf y}\,\bar\xi_\pm(x^0,{\bf y})\,R({\bf y},{\bf x})\,,
\ee
where $R$ depends on $x^0$ only implicitly through being a functional of $A_z$ on the constant-$x^0$ quantisation surface. To remove the unwanted vertices we need
$$
L_2[A]+L^{++-}[A]
=L_2[B]\,,\quad
L_f^{+-}[\bar\alpha,\,\alpha]+L^{++-}_f[A,\,\bar\alpha,\,\alpha]
=L_f^{+-}[\bar\xi,\,\xi]\,.
$$
These are satisfied by our previous solution for $B$ provided that $R$ satisfy
\newpage
$$
\left(\omega({\bf x})+\omega({\bf x}')\right)\,R({\bf x},\,{\bf x}')-
\int_\Sigma d^3{\bf y} \left(\omega({\bf y})\,A_z^P({\bf y})\right){\delta\over\delta
A_z^P({\bf y})}\,R({\bf x},\,{\bf x}')=
$$
$$
R({\bf x},\,{\bf x}')\,\partial_{\bar 0}^{-1}\partial_{\bar z}A_z({\bf x}')
-\left( \partial'_{\bar z}\partial^{\prime -1}_{\bar 0}\,R({\bf x},\,{\bf x}')\right)\,A_z({\bf x}')
$$
which can be solved in powers of $A_z$

$$
R({\bf x},\,{\bf x}')=\sum_{n=0}^\infty \int_{\Sigma} d^3{\bf y}_1..d^3{\bf y}_n \tilde\Gamma_n({\bf x},\,{\bf x}',\,{\bf y}_1..{\bf y}_n)\,A_z(x^0,\,{\bf y}_1)..A_z(x^0,\,{\bf y}_n)\,
$$
with the functions $\Gamma_n$ constructed iteratively from
$$
\tilde\Gamma_0({\bf x},\,{\bf x}')=\delta^3({\bf x}-{\bf x}')\,\one\,,
$$and
$$
\tilde\Gamma_n({\bf x},\,{\bf x}',\,{\bf y}_1..{\bf y}_n)=
$$
$$
 S\,{1\over \omega({\bf x})+\omega({\bf x}')+\omega({\bf y}_1)+..+\omega({\bf y}_n)}
\Big(\tilde\Gamma_{n-1}({\bf x},\,{\bf x}',{\bf y}_1..{\bf y}_{n-1})\partial_{\bar 0}^{-1}\partial_{\bar z}\delta({\bf y}_n-{\bf x}')$$
$$
-\delta({\bf y}_n-{\bf x}')\partial'_{\bar z}\partial^{\prime -1}_{\bar 0}\,\tilde\Gamma_{n-1}({\bf x},\,{\bf x}',{\bf y}_1..{\bf y}_{n-1})\Big)
$$
Having obtained the transformation the rest of the argument to identify the new
vertices as MHV amplitudes goes through
as before.

\section{Conclusions}
We have constructed a canonical transformation that takes the usual gauge theory action into one which generates the MHV rules. 
The use of light-front quantisation surfaces is a crucial first step.
It provides a natural interpretation of the off-shell continuation used in \cite{Witten} because the spinor $\lambda$ assocated to an off-shell momentum is the same as that associated with the on-shell momentum which has the same components within the quantisation surface.
Analyticity was also necessary to extract the off-shell vertices from the on-shell information contained in the MHV amplitudes.

Using these vertices and the scalar propagator we might begin to systematically construct the loop expansion in the usual way. 
To complete the task would require a choice of regulator
that preserved the structures we have exploited which appear to be
intrinsically four-dimensional. Also a complete treatment would require careful consideration of the singularities of the operator
$\partial_{\bar 0}^{-1}$ that is ubiquitous in our construction. The singularities are connected to the zero-modes generated by residual gauge
transformations and these have been thoroughly studied in the literature on light-front
quantisation.

We took a Lagrangian point of view, since this is the most familiar, but it is clear that the use of light-front quantisation surfaces is central
arguing that a Hamiltonian approach might be more natural.

\section{Acknowledgements}
It is a pleasure to acknowledge conversations with Bill Spence, Nigel Glover,
Valya Khoze and Anton Ilderton.

\appendix


\end{document}